\documentclass{article}
 \usepackage{sprocl}
 \usepackage{epsfig}
\newcommand{\zf}{{\tt{ZFITTER}}}
\newcommand{\flm}{\mu}
\newcommand{\mz}{M_{_Z}}
\newcommand{\afba}[1]{A^{#1}_{_{\rm FB}}}

\def\be{\begin{equation}}
\def\ee{\end{equation}}
\def\ba{\begin{eqnarray}}
\def\ea{\end{eqnarray}}

\begin{document}
\thispagestyle{empty}
\title{PROSPECTS AND PROBLEMS IN 
FERMION-PAIR PRODUCTION~\footnote{Presented by T. Riemann,
contribution to the Proceedings of the International Workshop {\em
Worldwide Study on Physics and Experiments with Future 
Linear $e^+e^-$ Colliders} (LCWS 99), Sitges (Barcelona), April 28 --
May 5, 1999   
}}
\author{M. JACK~\footnote{E--mail: Mark.Jack@ifh.de}, 
        T. RIEMANN~\footnote{E--mail: Tord.Riemann@ifh.de}}

\address{Deutsches Elektronen-Synchrotron DESY
\\
Platanenallee 6, D-15738 Zeuthen, Germany}


\maketitle\abstracts{
We discuss 2f production at LC energies ($f \neq t$).
This type of reaction has a big event number and may give interesting
hints to the existence and perhaps to details of New Physics like
susy, LQ, Z', etc.
For any search the radiative corrections have to be controlled carefully.
An interesting challenge for theoreticians could also be the Giga--Z
option of the TESLA project.  
}

\section{\label{introd}Introduction}
Fermion-pair production,
\ba
\label{ee2f}
e^+  + e^- \to f + {\bar f} + n \gamma, 
\ea
is among the scattering processes with the biggest expected counting rates at
a future Linear Collider (LC) running somewhere between 500 and 1000 GeV. 
This may be seen e.g. in Figure 1.(a) in \cite{Accomando:1997wt}.
If the Giga--Z option \cite{lcws99-moenig} of the TESLA--project
\cite{lcws99-brinkmann} will be realized, the $Z$--resonance region
could be additionally studied with a rate of about 
20$\times$10$^9$ $Z$--bosons.

The physics potential was summarized in several of the plenary talks 
\cite{lcws99plth-peskin}$^-$%
\nocite{lcws99pl-rueckl,lcws99pl-Tkaczyk}\cite{Richard:1999kq}
and quotations therein.
The main interest is concentrated on indirect high-energy searches with
reaction  
(\ref{ee2f}) for diverse phenomena beyond the Standard Model: 
$Z'$, $W'$, extra spatial dimensions, susy R-parity violation, leptoquarks,
preons, etc.

There is also some discovery potential of direct searches with the Giga--Z
option, e.g. for 
lepton number violation
\cite{frascati98-wilson}$^-$%
\nocite{oxford99-wilson}\cite{oxford99-TRiemann}.
But, the interest is focused here on precision studies of the 
Standard Model and its MSSM extension \cite{lcws99-heinemeyer}.

Whenever we have indirect searches, we
have to control the known physics extremely well.
Thus it might well be that we will have to face the challenge to provide
high precision predictions in two largely different situations -- at
the $Z$--peak and far away from it. 
Aiming at theoretical errors of {\em} single cross-section contributions of
one tenth of the experimental error, we need accuracies of
0.015\%  to 0.5\%,
depending on observables and rates; see Table 1 in \cite{Christova:1998tc}.

The success of indirect searches depends finally on the control of radiative
corrections to the reactions studied.
The cross-section for (\ref{ee2f}) is:
\ba
\label{sigma}
\sigma(s) \sim \int \frac{ds'}{s} \sigma^0(s') \rho(s'/s).
\ea
The radiative corrections may be roughly divided into two classes.
There are the {\em virtual corrections} to the basic, two-to-two-particle
scattering process $\sigma^0(s')$.
They are largely model-dependent and increase asymptotically at high
energies due to several mechanisms. 
A nice example for this are the one-loop t--quark corrections to $b$--pair
production shown in Figure 1 of \cite{Christova:1998tc}; see also
\cite{LKreptoLEP98}$^-$\nocite{Bardin:1999yd}\cite{Beccaria:1999xd}.
For a quantitative discussion of higher-order corrections see
e.g. \cite{Ciafaloni:1999uc}, with the estimate that uncertainties from them
at 1 TeV are not under 1\%.
Further, it has been stressed quite recently that virtual corrections of the
Standard Model may get completely modified in enlarged scenarios so that they
may lose their role as a reasonably estimated basis on which Born effects of
New Physics could appear \cite{Czakon:1999ha}.    
The other class of corrections are the {\em photonic} (and QCD--)
corrections $\rho(s'/s)$,
including real photon (and gluon) emission.
They are basically understood but have to be controlled carefully and
efficiently. 
A mini-review was given in \cite{lcws99-jadach}.
In the rest of this contribution, we will concentrate on a
semi-analytical approach advocated by the \zf\  team.
We aim at a realization of (\ref{sigma}) with {\em one-dimensional}
numerical integrations only, thus allowing for a flexible, accurate, but 
also very fast Fortran program well suited for the multi-parameter fits 
typically applied to the $Z$--resonance.

There have been performed countless improvements of the Fortran
packages for a description of the $Z$--lineshape, accompanied by
a number of dedicated numerical comparisons between different programs,
most recently e.g. in 
\cite{lcws99-jadach}$^-$\nocite{Bardin:1999gt}\cite{Christova:1999gh}. 
\section{\label{s-zf2}Photonic corrections at the $Z$--resonance} 
One of the open problems until recently was the accuracy of \zf\
at the $Z$--resonance with application of an acollinearity
cut, the so-called {\em realistic cut situation}.
The agreement e.g. with {\tt TOPAZ0} \cite{Montagna:1998kp} was
not considered to be satisfactory in \cite{Bardin:1999gt}.
After a complete recalculation it is excellent
now \cite{Christova:1999gh,Christova:1999cc}. 
This we may demonstrate with Table \ref{tab10acolifi2}.

\begin{table}[th]
\caption[]{
A comparison of predictions for muonic cross-sections and forward-backward
asymmetries around the $Z$--peak.
First row is without initial-final state interference, second row with,
third row the relative effect of that interference in per mil.
}
\begin{center}
\renewcommand{\arraystretch}{1.1}
\begin{tabular}{|c||c|c|c|c|c|}
\hline
\multicolumn{6}{|c|}{
{$\sigma_{\flm}\,$[nb] with $\theta_{\rm acol}<10^\circ$}}
\\ 
\hline
$\theta_{\rm acc} = 0^\circ$& $\mz - 3$ & $\mz - 1.8$ & $\mz$ & $\mz + 1.8$ &
$\mz + 3$  \\ 
\hline\hline
  & 0.21932  & 0.46287  & 1.44795  & 0.67725  & 0.39366 \\
{{\tt TOPAZ0}}  & 0.21776  & 0.46083  & 1.44785  & 0.67894  & 0.39491 \\
  & 
{\bf --7.16}    & 
{\bf --4.43}     &
{\bf --0.07}     &
{\bf +2.49}     &
{\bf +3.17}    
\\ 
\hline
  & 0.21928  & 0.46284  & 1.44780  & 0.67721  & 0.39360 \\
{{\tt ZFITTER}}  & 0.21772  & 0.46082  & 1.44776  & 0.67898  &
0.39489 \\ 
  &
{\bf --7.16}     &{\bf --4.40}     &{\bf --0.03 }    &{\bf +2.60}
&{\bf +3.27}   
\\ 
\hline 
\hline
\multicolumn{6}{|c|}{$\afba{\flm}$ with $\theta_{\rm acol}<10^\circ$} \\
\hline
$\theta_{\rm acc} = 0^\circ $& $\mz - 3$ & $\mz - 1.8$ & $\mz$ & $\mz + 1.8$ &
$\mz + 3$  \\ 
\hline\hline
  & --0.28450 & --0.16914  & 0.00033  & 0.11512  & 0.16107 \\
{{\tt TOPAZ0}}  & --0.28158 & --0.16665  & 0.00088  & 0.11385  &
0.15936 \\ 
  & 
{\bf +2.92}    & {\bf +2.49}     &
{{\bf +0.55}}     &{\bf --1.27}
&
{{\bf --1.71}}    \\ 
\hline
  & --0.28497 & --0.16936  & 0.00024  & 0.11496  & 0.16083 \\
{{\tt ZFITTER}}  & --0.28222 & --0.16710  & 0.00083  & 0.11392  & 0.15926 \\
  & {\bf +2.75}    & {\bf +2.27}     & {\bf +0.60}    &{\bf --1.03}
&{\bf --1.56}
\\
\hline 
\end{tabular}
\label{tab10acolifi2}
\end{center}
\end{table}

\section{\label{s-zf3}Photonic corrections above the $Z$--peak}
\begin{figure}[t] 
\begin{center}
\vspace*{-1.cm}
  \mbox{%
  \epsfig{file=%
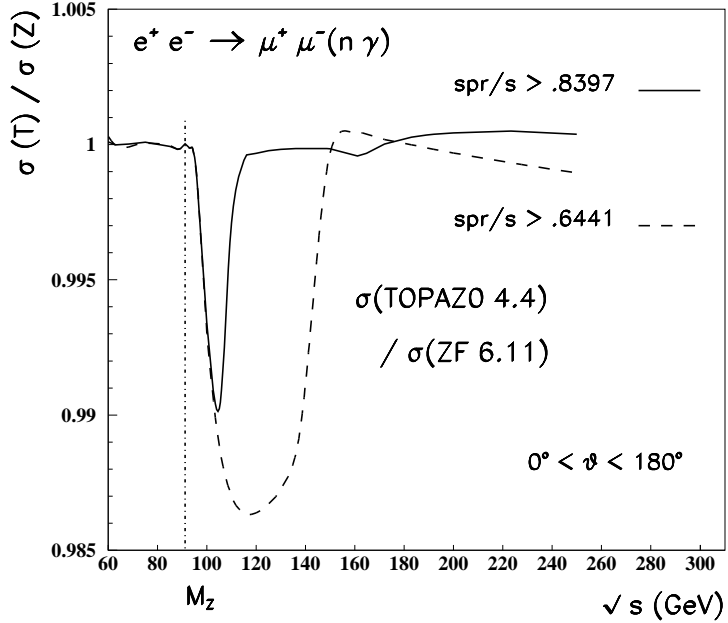
,width=10.cm   
         }}%
\vspace*{-0.3cm}
\caption[]
{Comparison of predictions for muon-pair production from  {\tt
ZFITTER} v.6.11  and {\tt TOPAZ0} v.4.4 with $s'$-cut. 
\label{compar-spr}}
\end{center}
\end{figure}

\begin{figure}[t] 
\begin{center}
\vspace*{-1.cm}
  \mbox{%
  \epsfig{file=%
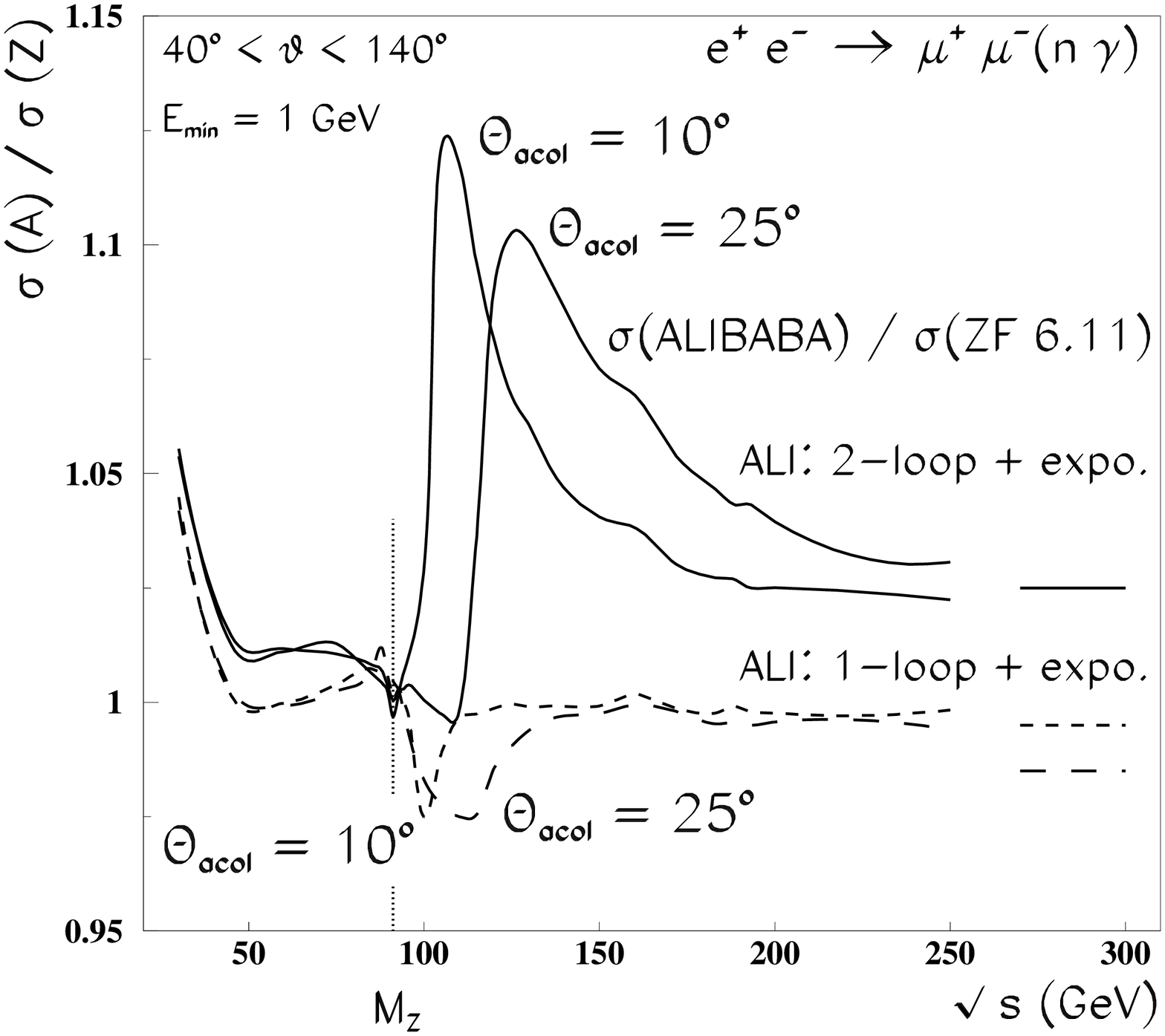,width=10.cm}}
\vspace*{-0.3cm}
\caption
{\sf
Muon-pair production cross-section ratios with 
$\theta_{\rm acol}^{\max}=10^{\circ}, 25^{\circ}$ and 
$\theta_{\rm acc}=40^{\circ}$ from
{\tt ALIBABA} v.2 (1990) and {\tt ZFITTER} v.6.11.
\label{compar-tzaz}
}
\end{center}
\end{figure}

At higher energies, the situation is much more involved since
radiative corrections generally will grow.
Further, there was a specific suppression of hard photonic
bremsstrahlung, and thus also of related classes of higher-order
corrections, in the vicinity of the $Z$--resonance. 
At an LC, we will have to control hard photonic corrections with great 
care, especially if the so-called radiative return is not
suppressed by dedicated cuts.
One may get an impression of all this from Figures 
\ref{compar-spr} and \ref{compar-tzaz}, 
where we show cross-section ratios from 
{\tt TOPAZ0} \cite{Montagna:1998kp} 
or 
{\tt ALIBABA} \cite{Beenakker:1991mb} 
with {\tt ZFITTER} \cite{Bardin:1999yd,home-ZFITTER}, with $s'$--cut
or with acollinearity cut. 
It may nicely be seen that for a sufficient suppression of the
radiative return to the $Z$--peak the programs agree well within the
expected errors, although it is also seen that higher order
corrections are treated differently.
For an acollinearity cut, allowing for more hard photonic corrections
than a correspondingly chosen $s'$--cut, the deviations between 
the programs get bigger.

\section*{Acknowledgments}
We would like to thank P. Christova and S. Riemann for a fruitful
cooperation when working on the problems discussed here.
We also would like to acknowledge the dedicated engagement of all the
members of the {\tt ZFITTER} team.
\section*{References}
\def\href#1#2{#2} 

\end{document}